\documentclass[conference]{IEEEtran}
\IEEEoverridecommandlockouts
\usepackage{cite}
\usepackage{amsmath,amssymb,amsfonts}
\usepackage{algorithmic}
\usepackage{textcomp}
\usepackage{xcolor}

\usepackage{bm}
\usepackage{array}
\usepackage{float}
\usepackage{caption}
\usepackage{subcaption}
\usepackage[colorlinks=true,bookmarks=false,citecolor=blue,urlcolor=blue]{hyperref} 

\usepackage{graphicx,calc}
\newlength\myheight
\newlength\mydepth
\settototalheight\myheight{Xygp}
\usepackage{comment}

\usepackage{amsthm,thmtools}

\declaretheoremstyle[
  headfont=\bfseries,
  bodyfont=\itshape,
]{myplain}
\declaretheoremstyle[
  headfont=\bfseries,
  bodyfont=\normalfont,
]{mydefinition}
\declaretheoremstyle[
  headfont=\itshape,
  bodyfont=\normalfont,
]{myremark}
\declaretheorem[
  style=myplain,
  name=Theorem,
]{theorem}

\def\BibTeX{{\rm B\kern-.05em{\sc i\kern-.025em b}\kern-.08em
    T\kern-.1667em\lower.7ex\hbox{E}\kern-.125emX}}

    \newcommand{\bpi}{\boldsymbol{\pi}} 

    \newcommand{\blam}{\boldsymbol{\lambda}}

\begin{document}

\title{Calculating the Capacity Region \\ of a Quantum Switch

\thanks{IT thanks NSF grant 1941583 and ORNL grant 4000178321 and KPS thanks NSF grant 2204985 and Cisco Systems.}
}

\author{\IEEEauthorblockN{Ian Tillman,$^{1,*}$
Thirupathaiah Vasantam,$^2$
Don Towsley,$^3$ and Kaushik P. Seshadreesan,$^4$}
\IEEEauthorblockA{
\\$^1$\textit{College of Optical Sciences}, 
\textit{University of Arizona}, Tucson, USA
\\$^2$\textit{Department of Computer Science}, \textit{Durham University}, Durham, UK
\\$^3$\textit{Manning College of Information \& Computer Sciences}, \textit{University of Massachusetts Amherst}, Amherst, USA
\\$^4$ 
\textit{School of Computing \& Information}, 
\textit{University of Pittsburgh}, Pittsburgh, USA
\\$^*$ijtillman@arizona.edu}}

\maketitle

\begin{abstract}
Quantum repeaters are necessary to fully realize the capabilities of the emerging quantum internet, especially applications involving distributing entanglement across long distances. 
A more general notion of this can be called a quantum switch, which connects to many users and can act as a repeater to create end-to-end entanglement between different subsets of these users.
Here we present a method of calculating the capacity region of both discrete- and continuous-variable quantum switches that in general support mixed-partite entanglement generation.
The method uses tools from convex analysis to generate the boundaries of the capacity region.
We show example calculations with illustrative topologies and perform simulations to support the analytical results.

\end{abstract}


\section{Introduction \label{sec:intro}}
A major step in the world of classical computing was the idea to design devices whose sole purpose was connecting many computing devices to each other~\cite{kleinrock2010early}.
This naturally progressed to entire infrastructures of such networking devices that now, thanks to research done on fiber optics, routers, and switching algorithms, allow for near instantaneous communication between computing devices thousands of kilometers apart \cite{hui2019introduction,bertsekas2021data}.
The internet is the backbone of many aspects of our day-to-day life, including secure online banking, social media, and remote education.
Quantum 2.0 technologies such as quantum computers, quantum sensor networks, and quantum simulators are in early stages of design and development, but in order to fully harness their capabilities we will need to dedicate ourselves to similar goals of interconnecting these devices and creating a quantum internet \cite{WEH18,dowlingMilburnQuantumRevolution,kimbleQuantumInternet}.
The fundamental resource that is created and shared in a quantum network is entanglement.
Some applications, quantum sensor networks in particular, will rely on many memories being able to generate entanglement between various subsets of themselves \cite{brady2022entangled,quantumSensingReview}.
This is analogous to a switch in classical computing, where one hub node connects to many end users and facilitates requests to send information between any specified set of users.
A quantum switch will rely on the users entangling themselves with the hub node first, followed up by the hub performing a `swap' operation that consumes the user-switch entanglements and creates the requested end-to-end entangled state between users \cite{diadamo2022packet,dai2021entanglement}.
Much like a classical switch, a quantum switch will be impacted by both the entanglement request rate from the users connected to it as well as the scheduling policy it adopts in order to service these requests.
However quantum switches will also be impacted by the often probabilistic nature of entanglement generation and manipulation, something that is effectively not present in classical networking.
One condition commonly imposed on switches, both classical and quantum, is the concept of \textit{stability} -- a term with many closely related definitions that all aim to give a quantitative way to test whether a switch is able to keep up with the incoming requests \cite{dai2021entanglement,vasantam2021stability}.

In concept, a quantum network can be just as general as a classical network---they are envisioned to facilitate the exchange of quantum information over local, metropolitan, or global scales \cite{chung2022design,clark2023entanglement,harney2022analytical,chen2023zero}. 
Quantum switches together with quantum repeaters are sufficient to enable arbitrary topologies for quantum networks. 
The bulk of what these quantum networks will be useful for is distributing entanglement between quantum memories, as most quantum-enhanced activities are based on utilizing the non-classical correlations that come from quantum entanglement.
Devices and protocols have been theorized and tested for many different platforms of entanglement, many of which are designed with just one particular use case in mind.
However there are still many fundamental, platform and application independent questions to be answered in this field, and so there are many groups that study quantum networks from as general of a perspective as they can \cite{cicconetti2022resource}.
Many of these questions deal with bounds on the \textit{capacity} of a quantum channel, a term which normally means the maximum rates at which a network can create entanglement.
For networks which can create many different entanglements this is extended to a \textit{capacity region}, where we consider what possible combinations of rates are possible.
In these scenarios it is often unclear how to define the 'optimal' rate vector, so in general it is often more useful to characterize the entire region and then let the application decide what vector within the region best suites its needs.
Some work has been done to characterize the capacity region of quantum switches~\cite{thiru,VGNT21,vardoyan2023capacity,promponas2023full,tillman2022continuous}, however these characterizations are often not conducive to certain calculations.
For example, we cannot easily verify whether a given rate vector is in the capacity region or not.
Motivated by this, in the present work we showcase a method to analytically calculate the capacity region of a single quantum switch that is allowed to use general swap operations to create entanglement between any subset of the many users connected to it.
We use elementary convex analysis to derive our algorithm and then present example calculations alongside simulation data that supports our analytic results.

The manuscript is organized as follows: We start with some background on quantum repeaters, quantum switches, and convex analysis in Section \ref{sec:background}.
In Section \ref{sec:model} we describe our model for a quantum switch, discuss simulating the stability of a quantum switch, and explain the complications that arise when using continuous variable encodings.
In Section \ref{sec:results} we describe the algorithm to generate the capacity region of a quantum switch.
Section \ref{sec:exampleCalculations} contains example calculations and discussions on how to interpret our results.
Finally, in Section \ref{sec:discussion}, we bring up a particular problem that can be solved faster than expected and then discuss the complexity of our algorithm

\section{Background \label{sec:background}}
\subsection{Related Work}
Much focus of quantum networking research today is spent on quantum repeaters, devices that are used to extend the range of quantum entanglement distribution.
In classical optical networking one can place a repeater node at any point along a long link to read in a lossy, noisy signal and send a restored copy.
This alleviates the user on the other end of having to deal with noise, and in some cases the signal would be unreadable at the other end without the repeater.
When a repeater can successfully restore the signal, multiple repeaters can be chained to extend a channel's usable distance even more.
This is, for example, how we are able to communicate reliably with fiber optic cables under the oceans even though we expect roughly $0.2\ \mathrm{dB/km} \times 5500\ \mathrm{km} = 1100\ \mathrm{dB}$ of loss in optical power between New York City and London.
Unfortunately a quantum analog of the classical optical repeater is impossible for quantum communication channels because of the No-Cloning theorem~\cite{noCloning}.
The theorem states that it is impossible to make a copy of an arbitrary quantum state, which is exactly what a repeater would need to do.
Because of this, researchers have looked for ways to effectively amplify quantum signals that get around this.

It has been proven~\cite{PLOB} that there is a bound on the generation rate, $R$, of quantum entanglement that can be shared between two users separated by a channel with transmissivity $\eta$ that does not contain a repeater:
\begin{equation}
    R \leq C(\eta) = -\log_2(1-\eta)\ \ \mathrm{ebits/mode}
\end{equation}
\noindent where $C$ is the channel capacity and 'ebit' stands for entangled bit.
Note that as $\eta \rightarrow 0$ we have $C(\eta) \sim \eta$.
However, some repeater designs have been shown to support capacities that scale as $\sqrt{\eta}$, the half-channel loss, when the repeater is placed between Alice and Bob.
Often these repeater protocols perform poorly when $\eta \approx 1$ due to inherently probabilistic operations that can be avoided with direct transmission.
However these protocols are designed to exhibit a slower drop in capacity as transmissivity decreases, so there will be a critical transmissivity (i.e. channel distance) where the repeater protocol is able to generate more ebits per mode than a direct transmission protocol.

Many different approaches have been taken to implement repeaters~\cite{munro2015inside,repeaterGenerations,yan2021survey,azuma2022quantum}, however outside of one specific architecture that requires special care in our model we will not discuss how repeaters have been modelled or implemented.
The one repeater design we will briefly look at uses the fact that the No-Cloning theorem only disallows deterministic amplification and says nothing about probabilistic protocols.
One class of probabilistic methods is known as noiseless linear amplification (NLA) which seeks to probabilistically amplify an incoming quantum-limited state without amplifying its noise~\cite{ralph2009nondeterministic,he2021noiseless}.
Unfortunately it has been proven that ideal NLA must have a probability of success equal to zero, so the best one can hope to do is approximate the action of NLA~\cite{PJCC13}.
One example of this is the Quantum Scissor (QS)~\cite{peggQuantumScissor,diasRalph2018,kaushikPRR} which approximates NLA by performing projection measurements.
The QS is the basis of the continuous variable repeater model used in this paper, which we describe in detail in Section \ref{subsec:CV}.

\subsection{Quantum Switches}
\label{subsec:quantumswitches}
A natural extension to the idea of a quantum repeater is a quantum switch, which performs the same operation as a repeater but there are many users connected to it instead of just two.
This is analogous to a classical switch or router, where many users connect to one device that can route information between any pair of them.
In the quantum scenario, one of the models of a quantum switch considers end users making requests to generate end-to-end entanglement rather than simply routing a packet of data between users as in a classical scenario. 
Because the switch is responsible for generating entanglement between two end users, the physical model or implementation of a specific quantum switch will differ for different entanglement platforms.
For example a request for a Bell state can be serviced with atomic ensembles~\cite{DLCZ} while a request for a continuous variable entanglement may require devices like two-mode squeezed vacuum (TMSV) generators, homodyne detection, and distillation schemes~\cite{tillman2022supporting,diasCVRepeater}.

Regardless of the underlying model, we can reduce our analysis to a set of users and a set of end-to-end states that subsets of users can request.
We call the entanglements between the users and the switch \textit{elementary entanglements}, and we call the operations that consume elementary entanglements to create end-to-end states \textit{swap operations}.
We refer to the switch creating an end-to-end entanglement as the switch servicing an entanglement flow, with each possible end-to-end state being its own flow.
We then use two sets of parameters: the probabilities that a given user generates an elementary entanglement with the switch, and the probability that the switch successfully creates a given end-to-end entanglement provided that all necessary elementary entanglements exist.
With these parameters we can analyze the performance of a quantum switch, such as the end-to-end state generation rates it can support, while being agnostic to its underlying physical platform.
Some work has been done to characterize these models, mostly concerning the scheduling policies one should use to handle competing requests~\cite{thiru,dai2021entanglement}.
These papers also characterize the set of request rates that are supportable by the switch, a set called the \textit{Capacity Region}, but none of them have provided a direct method for calculating the boundaries of this set.
Knowing the boundaries would, for example, make it easy to determine whether or not a specific request rate vector will be supportable by the switch.
Two current methods for solving this problem involve either partitioning~\cite{thiru} or optimization~\cite{panigrahy2023capacity}, neither of which can be done efficiently.

\subsection{Extreme Point Analysis \label{sec:extremepoint}}
Our method of calculating the capacity region of a quantum switch uses ideas from convex analysis, so we introduce the necessary tools here.
A set $S$ is said to be \textit{convex} if $x_1,x_2 \in S \Rightarrow tx_1 + (1-t)x_2 \in S$ for all $0< t < 1$.
A convex combination of points $x_1,\cdots,x_n$ is expressed as $\sum_{k=1}^n \alpha_k x_k$ where $\alpha_k > 0$ and $\sum_{k=1}^n \alpha_k = 1$.
We can further define $\mathrm{cch}(S)$ to be the set of all convex combinations of elements of $S$, known as the closed convex hull of $S$.
A point $e\in S$ is said to be an \textit{extreme point} of $S$ if it cannot be expressed as a convex combination of two or more unique points in $S$, i.e. $e = \sum_{k=1}^n \alpha_k x_k \Rightarrow x_1 = x_2 = \cdots = x_n$.
Define $\mathcal{E}(S)$ to be the set of extreme points of $S$.
The Krein-Milman Theorem says that a compact convex set can be fully recreated by the closed convex hull of its extreme points~\cite{simon2011convexity}, $S = \mathrm{cch}\left(\mathcal{E}(S)\right)$ for any compact convex set $S$.
This implies that any point in $S$ can be expressed as a convex combination of extreme points.
The following theorem will be useful in later sections:
\begin{theorem}
Let $L:S\rightarrow S^\prime$ be a linear transformation acting on a compact convex $S$, then $\mathcal{E}\left( L(S) \right) \subseteq L\left( \mathcal{E}(S)\right)$
\end{theorem}
In other words, if we know all extreme points of $S$ then we have a way of generating a set containing all extreme points of $L(S)$.
For completeness, we include a proof of this fact here.
\begin{proof}
Start with an extreme point $e^\prime \in \mathcal{E}\left(L(S)\right)$ and write it as a convex combination of transformed extreme points of $S$.
Say that $L(x_0)=e^\prime$ and $x_0 = \sum_{k=1}^n \alpha_k e_k$, where $e_k\in \mathcal{E}(S)$ for $1\leq k \leq n$ and $\sum_{k=1}^n \alpha_k = 1$.
\begin{equation}
    e^\prime = L \left( \sum_{k=1}^n \alpha_k e_k \right) = \sum_{k=1}^n \alpha_k L(e_k)
\end{equation}
Because $L(e_k)\in L(S)$, we have written an extreme point of $L(S)$ as a convex combination of points in $L(S)$. This means that $L(e_1)=L(e_2)=\cdots = L(e_n)$, implying that we can replace $L(e_k)$ with $L(e_1)$ for $2\leq k\leq n$:
\begin{equation}
    e^\prime = \sum_{k=1}^n \alpha_k L(e_1) = L(e_1)
\end{equation}
proving that $e^\prime \in L\left( \mathcal{E}(S)\right)$.
\end{proof}

If $L(S)$ is a subset of $\mathbb{R}^n$ and has finitely many extreme points we can use an algorithm like QuickHull~\cite{quickhull} to reduce $L\left( \mathcal{E}(S)\right)$ to $\mathcal{E}\left(L(S)\right)$, stripping away all non-extreme points and allowing us to easily define the planar/hyperplanar boundaries of $L(S)$ using an algorithm like Delaunay Triangulation~\cite{preparata2012computational}.

\section{Model \label{sec:model}}
\subsection{Quantum Switch Model \label{subsec:switchModel}}

\begin{figure}
    \centering
    \includegraphics[width=0.3\textwidth]{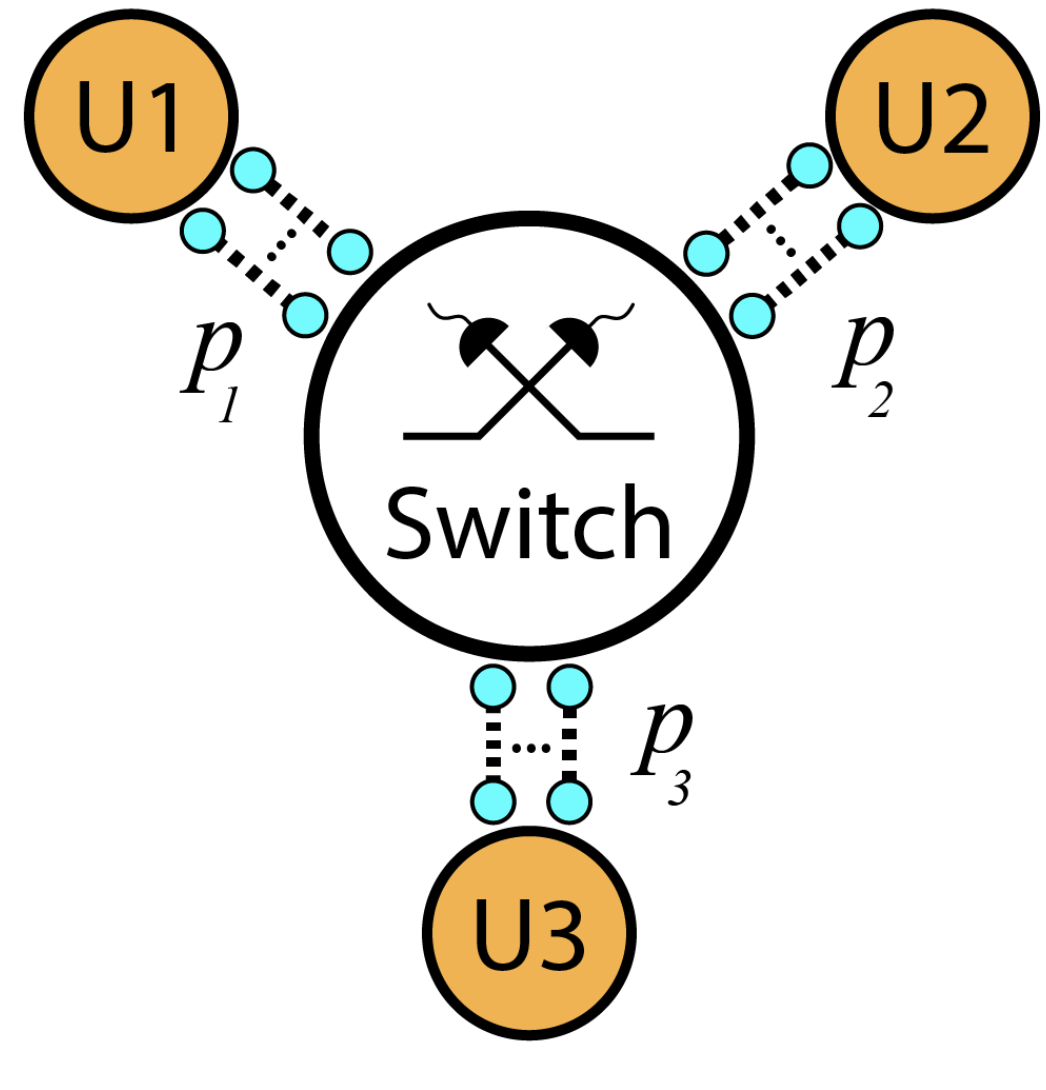}
    \caption{An example of the underlying Physical topology we consider. On top of this is a virtual topology containing the $M$ different flows the switch is capable of generating.}
    \label{fig:physicalTopology}
\end{figure}

Our quantum switch model is based on a hub-and-spoke topology, with the switch acting as the central hub node and each end users' one or many memories being the spokes.
Let $K$ be the number of end users, $D_i$ be the number of memories at user $i$, $\mathbf{D}$ be a $K\times 1$ vector whose $i$th component is $D_i$, and $M$ be the number of entanglement flows the switch supports.
$D_i$ can be thought of as the degree of spatial multiplexing that user $i$ implements.
Associated with each user memory there is a memory at the switch.
Fig. \ref{fig:physicalTopology} shows an example of this physical topology we consider.
Let $\mathbf{M}=\left[ \mathbf{M}_{i,j} \right]$, where $\mathbf{M}_{i,j}$ is the number of user $i$ elementary links needed to service flow $j$.
During each timestep the switch and users perform the following actions:

\begin{enumerate}
    \item Each unoccupied user memory attempts to generate entanglement with its associated switch memory
    \item The switch uses the information of which elementary links currently exist to make a scheduling decision
    \item The switch services the flow(s) it chose to serve by attempting entanglement swap operations
    \item The switch discards all elementary entanglements that are older than their maximum allowed lifetime.
\end{enumerate}

\noindent User $i$'s memories each have a probability $p_i$ of successfully entangling with the switch, and the swap success probability for flow $j$ is $q_j$.
User $i$'s elementary links can be stored for $\tau_i$ timesteps, where the timestep it is created counts as the first timestep (i.e. $\tau=1$ means elementary links are either used immediately or discarded).
This paper only considers \textit{bufferless} switches, defined as switches where all memories have $\tau=1$.
We do not take into account any measure of quality of the end-to-end entanglements generated.

Define an \textit{arrangement}, $\mathbf{a}(n)$, $n\in \{1,2,\cdots\}$ to be a $K\times 1$ vector whose $i$th component is the number of elementary links between user $i$ and the switch available in timestep $n$.
Here $\mathbf{a}(n) \in \mathcal{A}$, where $\mathcal{A}$ is the set of all possible elementary link arrangements.
Give $\mathcal{A}$ an ordering and call the $k$th arrangement $\mathbf{a}^{(k)}$.
It is important to note that we do not assume any particular model for the memories, swap operations, or entangled states in our model.
To apply our analysis to specific physical models one would need to calculate the probability parameters required for our methods and then follow the steps outlined in the rest of the paper.

We allow the switch to effectively serve multiple swaps simultaneously, given that there are enough existing elementary links to service all of them.
To generalize the idea of the switch choosing a flow to service, we define a matching $\bpi$ to be an $M\times 1$ vector where the $j$th element is the number of requests of flow $j$ that the switch services in a particular timestep.
Therefore $\mathbf{M}\bpi$ is a vector that contains the number of elementary links required from each user to service the matching $\bpi$.
A matching is \textit{valid} if there are enough memories available to support all the requests it asks for, i.e. $\mathbf{M}\bpi \leq \mathbf{a}$.
Define $\mathcal{M}$ to be the set of all matchings that would be valid if every elementary link existed simultaneously, $\mathcal{M} = \{ \bpi\ |\ \mathbf{M}\bpi \leq \mathbf{D} \}$.
Give this set an ordering and call the $k$th matching $\bpi^{(k)}$.
Let $\blam^*=(\lambda^*_1,\cdots,\lambda^*_M)^T$ denote a servicing rate vector, where $\lambda^*_j$ is the average number of end-to-end entangled states of flow $j$ generated per timestep.
Similarly, let $\blam=(\lambda_1,\cdots,\lambda_M)^T$ denote a request rate vector, where $\lambda_j$ is the average number of end-to-end entangled states of flow $j$ requested by the users per timestep.
We say the switch is stable if it can guarantee that the average number of timesteps it takes for the switch to service a specific request is always finite~\cite{thiru}.
Clearly if the switch is stable then $\boldsymbol{\lambda}^* = \boldsymbol{\lambda}$, though the converse is not necessarily true.

A switch scheduling policy is defined as the set of probabilities that matching $\bpi$ is selected given that elementary link arrangement $\mathbf{a}$ exists,
\begin{equation*}
\mathbb{P}\left(\bpi\ \mathrm{chosen}\ |\ \mathbf{a}\right) \equiv c_{\bpi,\mathbf{a}},\ \ \bpi \in \mathcal{M},\ \mathbf{a}\in \mathcal{A}.
\end{equation*}
If $\mathbf{M}\bpi \nleq \mathbf{a}$ then we set $c_{\bpi,\mathbf{a}}=0$ since that means matching $\bpi$ cannot be serviced when $\mathbf{a}$ is the elementary link arrangement.
We store these probabilities in a matrix in the following way:
$$ \mathbf{c}= 
\begin{bmatrix}
c_{\bpi^{(1)}, \mathbf{a}^{(1)}} & c_{\bpi^{(1)}, \mathbf{a}^{(2)}} & \ldots\\
c_{\bpi^{(2)}, \mathbf{a}^{(1)}} & c_{\bpi^{(2)}, \mathbf{a}^{(2)}} & \ldots\\
\vdots & \vdots & \ddots\\
\end{bmatrix}
\overset{\Delta}{=} 
\begin{bmatrix}
c_{1,1} & c_{1,2} & \ldots\\
c_{2,1} & c_{2,2} & \ldots\\
\vdots & \vdots & \ddots\\
\end{bmatrix}.
$$
$\mathbf{c}$ is an $ \left|\mathcal{M}\right|\times \left|\mathcal{A}\right| $ matrix.
Because the terms of $\mathbf{c}$ are probabilities and the switch can only choose one matching at a time, we include the following two constraints:
\begin{gather}
\label{eqn:cCondition}
c_{i,j} \geq 0 \ \ \ \forall i,j \\ \sum_i c_{i,j}= 1\ \ \ \forall j \nonumber
\end{gather}
where the last constraint is equality because we allow the matching $\bpi=[0,\cdots,0]^T$, therefore even the switch doing nothing is considered a scheduling decision.

Having defined a scheduling policy in terms of matching probabilities conditioned on elementary link arrangements, to get the average number of end-to-end entangled states generated per timestep we need to determine the distribution of those arrangements.
Because we only consider the case $\tau=1$, arrangements are independently and identically distributed across timesteps.
Denote the elementary link arrangement distribution $\mathbf{P} = \left[ \mathbb{P}(\mathbf{a}^{(1)}), \mathbb{P}(\mathbf{a}^{(2)}),\cdots, \mathbb{P}(\mathbf{a}^{({|\mathcal{A}|})}) \right]^T$ that can be used for every timestep.
Let $\bpi^*$ denote a matrix whose $j$th column is $\bpi^{(j)}$ and $\mathbf{Q}=\mathrm{diag}\left( q_1,\cdots,q_M \right)$.
Theorem 4.1 in \cite{thiru} says that any servicing rate vector in $\Lambda^*$ can be expressed as
\begin{equation*}
    \blam^* = \sum_{\mathbf{a}\in \mathcal{A}} \mathbb{P}(\mathbf{a}) \sum_{\bpi \in \mathcal{M}} c_{\bpi,\mathbf{a}} \mathbf{Q}\bpi
\end{equation*}
which is equivalent to
\begin{equation}
\label{eqn:servicingRate}
    \blam^* = \mathbf{Q}\bpi^*\mathbf{cP}.
\end{equation}
This can be understood as a weighted average of matching vectors that is then scaled anisotropically by each flows' associated swap probability of success.

The main goal of this paper is to calculate the set of possible servicing rate vectors, which is a set over all valid scheduling policies:
$$ \Lambda^* \equiv \{ \blam^*\ |\ \mathbf{c}\ \mathrm{valid} \}. $$
Although this is a full description of the servicing region, it does not provide a way of calculating it.
In Section \ref{sec:results} we present a method that uses (\ref{eqn:servicingRate}) to provide a better characterization of $\Lambda^*$. 

One distinction still needs to be made; we define the \textit{capacity region} of a switch to be the set $\Lambda$ of request rate vectors $\blam$ that the switch can service while remaining stable.
This corresponds to $\Lambda^*$ once the boundary of $\Lambda^*$ is removed, i.e. $\Lambda = \Lambda^* \setminus \partial\Lambda^*$~\cite{thiru}.
In other words, if it is possible for the switch to create end-to-end entangled states with rate vector $\blam^*$ then there exists a scheduling policy that can support a request rate vector $\blam < \blam^*$.
Because this distinction is trivial to take into account and has no effect on physical implementations, we ignore it and focus solely on the easier to manage servicing region for the remainder of the paper.

\subsection{Simulating a Quantum Switch}
\label{subsec:simulation}
As mentioned in Section \ref{subsec:quantumswitches} some prior work has studied how to efficiently allocate the switch's resources when different flows compete with each other.
For our model, the most useful result is a proven throughput optimal scheduling policy~\cite{thiru} that a switch can adopt to stably service any incoming request rate vector $\blam$ that is in the capacity region $\Lambda$ (i.e. the interior of the servicing region $\Lambda^*$).
This policy is based on the MaxWeight framework, and it selects the following matching in timestep $n$:
\begin{equation}
    \underset{\mathbf{M}\bpi \leq \mathbf{a}(n)}{\mathrm{arg}\ \mathrm{max}} \sum_{i=1}^M Q_i(n) q_i \pi_i
\end{equation}
where $Q_i(n)$ is the number of requests of flow $i$ in the queue at timestep $n$, $\pi_i$ is the $i$th element of $\bpi$, and we maximize over all matchings $\bpi$ that satisfy $\mathbf{M}\bpi \leq \mathbf{a}(n)$.
Because this scheduling policy is throughput optimal for the scenario we are studying, we can use it to simulate a switch's performance after calculating its capacity region.
We will use these Monte Carlo simulations to corroborate our analytical calculations of servicing regions.
Details on this are at the beginning of Section \ref{sec:exampleCalculations}.

\subsection{Continuous Variables \label{subsec:CV}}
Later we will address how the use of continuous variable (CV) protocols will affect the quantum switch model, but first we describe the particular repeater architecture that we base it on.
The basis for our CV repeater model is previous work~\cite{furrerAndMunro,kaushikPRR,diasCVRepeater,tillman2022supporting} that studied a quantum scissor (QS) based repeater, which uses two-mode squeezed vacuum (TMSV) states as sources.
This repeater provided an inherent directionality, i.e. \textit{parity}, to each half-channel and the final end-to-end state can only be created from two half-channels of opposite parity.
We assume that every elementary link is capable of being either parity and for simplification we give each memory a $\frac{1}{2}$ probability of being either.
This may seem inappropriate because the protocol to create the links requires that the parity be agreed upon beforehand, but this repeater design heavily relies on multiplexing so alternating attempts is very natural.
This complicates our model from the last section slightly, as we now have to keep track of which parity each elementary link is.
We will address this issue by splitting one elementary link into two virtual links, one for each parity, and altering the probabilities $\mathbb{P}(\mathbf{a}^{(j)})$ to enforce the fact that both parities cannot exist in the same elementary link simultaneously.
This in turn creates new virtual flows, since we can create entanglement between the same two users with two different parity combinations of virtual elementary links.
This model imposes the restriction that our switch can only create bipartite end-to-end states, as the QS-based repeater does not trivially generalize to multipartite entanglement generation.
Therefore if we are given a switch topology that creates $M$ different bipartite CV end-to-end states, we can use a non-CV model with $2M$ flows to calculate the capacity region.
If one wants to treat the two end-to-end parities the same then they can simply reduce all derived bounds by combining the two equivalent virtual flows associated with each real flow; however in practice it may be useful to keep track of the different parities.

\section{Results \label{sec:results}}
Though we can input any distribution of elementary link arrangements, if we treat the generation of all link-level entanglements as independent Bernoulli trials then in the DV case we have
$$ \mathbb{P} \left(\mathbf{a}(n)=\mathbf{a} \right) = \mathbb{P}(\mathbf{a}) = \prod_{i=1}^K \genfrac(){0pt}{0}{D_i}{a_i} p_i^{a_i}(1-p_i)^{(D_i - a_i)},\ \mathbf{a}\in \mathcal{A} $$
Observing that the set of valid $\mathbf{c}$ matrices, $C$, is convex and that the relationship between a scheduling matrix $\mathbf{c}$ and the servicing vector $\blam$ shown in (\ref{eqn:servicingRate}) is a linear transformation, if we can find all extreme points of $C$ then we have a way to calculate all extreme points of $\Lambda^*$.
In order for this computation to be feasible we need to find a systematic way to generate all extreme points of $C$ and show that it will finish in a finite amount of time.

We first characterize the extreme points of $C$.
In short, every column of an extreme point of $C$ must contain a single $1$ and the other entries must be $0$.
To see why this must be true, we will prove that any matrix that does not have this property cannot be an extreme point of $C$.
Let $\mathbf{c}_0\in C$ be a matrix whose $j$th column has exactly two nonzero elements, $c_{i_1,j}=\alpha_1$ and $c_{i_2,j}=\alpha_2$, meaning $\alpha_1 + \alpha_2=1$.
Let $\mathbf{c}_1$ and $\mathbf{c}_2$ be matrices with entries identical to $\mathbf{c}_0$ except in $\mathbf{c}_1$ we have $c_{i_1,j}=1$ and $c_{i_2,j}=0$, while in $\mathbf{c}_2$ we have $c_{i_1,j}=0$ and $c_{i_2,j}=1$.
Clearly if $\mathbf{c}_0\in C$ then $\mathbf{c}_1,\mathbf{c}_2 \in C$ as well. 
Because $\mathbf{c}_0$ can be written as a convex combination of these two:
$$ \mathbf{c}_0 = \alpha_1 \mathbf{c}_1 + \alpha_2 \mathbf{c}_2 $$
$\mathbf{c}_0$ cannot be an extreme point of $C$.
This argument can be trivially extended to cases where more than two elements in a column are nonzero, showing that all extreme points of $C$ cannot have more than one nonzero entry per column.
These matrices being column stochastic implies that this single nonzero value must be $1$ in every column.
Because these matrices are finite in size there will be a finite number of matrices with this property and thus a finite number of extreme points of $C$, meaning there will be a finite number of extreme points of the servicing region $\Lambda^* = \mathbf{Q}\bpi^*C\mathbf{P} \subseteq \mathbb{R}^M$ and we have a method of generating a subset of $\Lambda^*$ that contains them all.

Once we filter the set $\mathbf{Q}\bpi^* \left[\mathcal{E}(C)\right]\mathbf{P}$ down to just $\mathcal{E}\left(\Lambda^*\right)$ using an algorithm like QuickHull, we can characterize the set much easier as $\Lambda^* = \mathrm{cch}\left(\mathcal{E}\left(\Lambda^*\right)\right)$.
There also exist algorithms like Delaunay Triangulation~\cite{preparata2012computational} to convert the extreme points of $ \Lambda^* $ to the hyperplanar boundaries of $\Lambda^*$.
We know that the boundaries of $\Lambda^*$ will be hyperplanar because it is a convex compact subset of $\mathbb{R}^M$ with a finite number of extreme points, meaning the region must be a polytope~\cite{simon2011convexity}.
Thus all its bounds will be of the form
$$ \sum_{j=1}^M \alpha_j \lambda^*_j \leq \gamma_{\boldsymbol{\alpha}} $$
where $\boldsymbol{\alpha}$ is an $M\times 1$ vector of weights and $\gamma_{\boldsymbol{\alpha}} \in \mathbb{R}$ is the upper bound of the weighted sum of rates.
These can be used to quickly test whether a given rate vector $\blam^*$ is an element of the servicing region or not.
Without these boundaries one would need to find a valid scheduling matrix, $\mathbf{c}$, that gives $\blam^*$ using (\ref{eqn:servicingRate}), which is possible but very challenging.
It is often more convenient to write these bounds in the form:
$$ \sum_{j=1}^M \alpha_j \frac{\lambda^*_j}{q_j} \leq \beta_{\boldsymbol{\alpha}} $$
where the ratio $\lambda_j^*/q_j$ can be thought of as an effective rate because $1/q_j$ is the average number of attempts it takes for the swap operation of flow $j$ to succeed and service one request.

\section{Example Calculations}
\label{sec:exampleCalculations}
As mentioned in Section \ref{subsec:simulation}, we use simulation data to support our analytical results.
We mark a request rate vector that was simulated to be stable by placing a blue dot at its corresponding point in space, where we consider a vector to be stable if we simulate the system with that request arrival rate for $10^6$ timesteps and a linear regression on the function $\frac{1}{M}\sum_{i=1}^M Q_i(n)$ returns a slope of less than $10^{-4}$ $\mathrm{requests}/\mathrm{timestep}$.
This threshold slope was chosen somewhat arbitrarily, but the transition from stable to unstable is very sharp and this value seems to consistently cull clearly unstable vectors while correctly labelling clearly stable vectors.
We see in the figures below that there are many points being labelled as stable that are outside the planar boundaries that we predict bound the capacity region.
This is to be expected since these data are numerical in nature, and points just outside of the capacity region are expected to result in very small slopes anyway.
In fact, a plane having an inconsistent pattern of these points peeking through can be used as evidence that the bound is correct.
A plane with too low of a bound will underestimate the capacity region (i.e. the bound will be too tight) and have stable points consistently appearing outside the boundary it forms.
Similarly, planes with too high of a bound will overestimate the capacity region (i.e. the bound will be loose) and have no points coming through.
Furthermore, a plane at an incorrect angle will exhibit a different density of these incorrectly labelled points at different parts of the plane, so a homogeneous pattern of these can tell us that the plane is at the correct height and angle.
\subsection{Triangular Topology}
\label{subsec:triangle1}
We start with a triangular topology with probabilities $p=q=\frac{1}{2}$:
\begin{table}[ht]
    \normalsize
    \centering
    \setlength\extrarowheight{3pt}
    \begin{tabular}{c|c}
         Flows: & \{(1,2), (2,3), (1,3)\}  \\
         $p$ & $\{ \frac{1}{2}, \frac{1}{2}, \frac{1}{2} \}$ \\
         $q$ & $\{ \frac{1}{2}, \frac{1}{2}, \frac{1}{2} \}$ \\
         $D$ & $\{1, 1, 1\}$
    \end{tabular}
\end{table}

\noindent thus we have $\mathbf{P}=[\frac{1}{8},\cdots,\frac{1}{8}]^T$.
The $\mathbf{M}$ matrix is easily generated from this information:
$$
\mathbf{M}
=
\begin{bmatrix}
1 & 0 & 1 \\
1 & 1 & 0 \\
0 & 1 & 1 \\
\end{bmatrix}.
$$
No two flows can be serviced simultaneously, so $\bpi^*$ also has a very basic structure:
$$
\bpi^*
=
\begin{bmatrix}
0 & 1 & 0 & 0 \\
0 & 0 & 1 & 0 \\
0 & 0 & 0 & 1 \\
\end{bmatrix}.
$$
Defining $\mathbf{a}^*$ to be a matrix whose $i$th column is $\mathbf{a}^{(i)}$, we can also define our ordering of elementary link arrangements as:
$$
\mathbf{a}^*
=
\begin{bmatrix}
0 & 0 & 0 & 0 & 1 & 1 & 1 & 1 \\
0 & 0 & 1 & 1 & 0 & 0 & 1 & 1 \\
0 & 1 & 0 & 1 & 0 & 1 & 0 & 1 \\
\end{bmatrix}.
$$
From these orderings we can fill in the $\mathbf{c}$ matrix with its zero and non-zero terms:
$$
\mathbf{c}
=
\begin{bmatrix}
c_{1,1} & c_{1,2} & c_{1,3} & c_{1,4} & c_{1,5} & c_{1,6} & c_{1,7} & c_{1,8} \\
0 & 0 & 0 & 0 & 0 & 0 & c_{2,7} & c_{2,8} \\
0 & 0 & 0 & c_{3,4} & 0 & 0 & 0 & c_{3,8} \\
0 & 0 & 0 & 0 & 0 & c_{4,6} & 0 & c_{4,8} \\
\end{bmatrix}.
$$
For example, $c_{4,7}$ must be zero because the $7$th column of $\mathbf{a}^*$ is $[1,1,0]^T$ which cannot support the matching in the $4$th column of $\bpi^*$, $[0,0,1]^T$.

\begin{figure}
    \centering
    \includegraphics[width=0.45\textwidth]{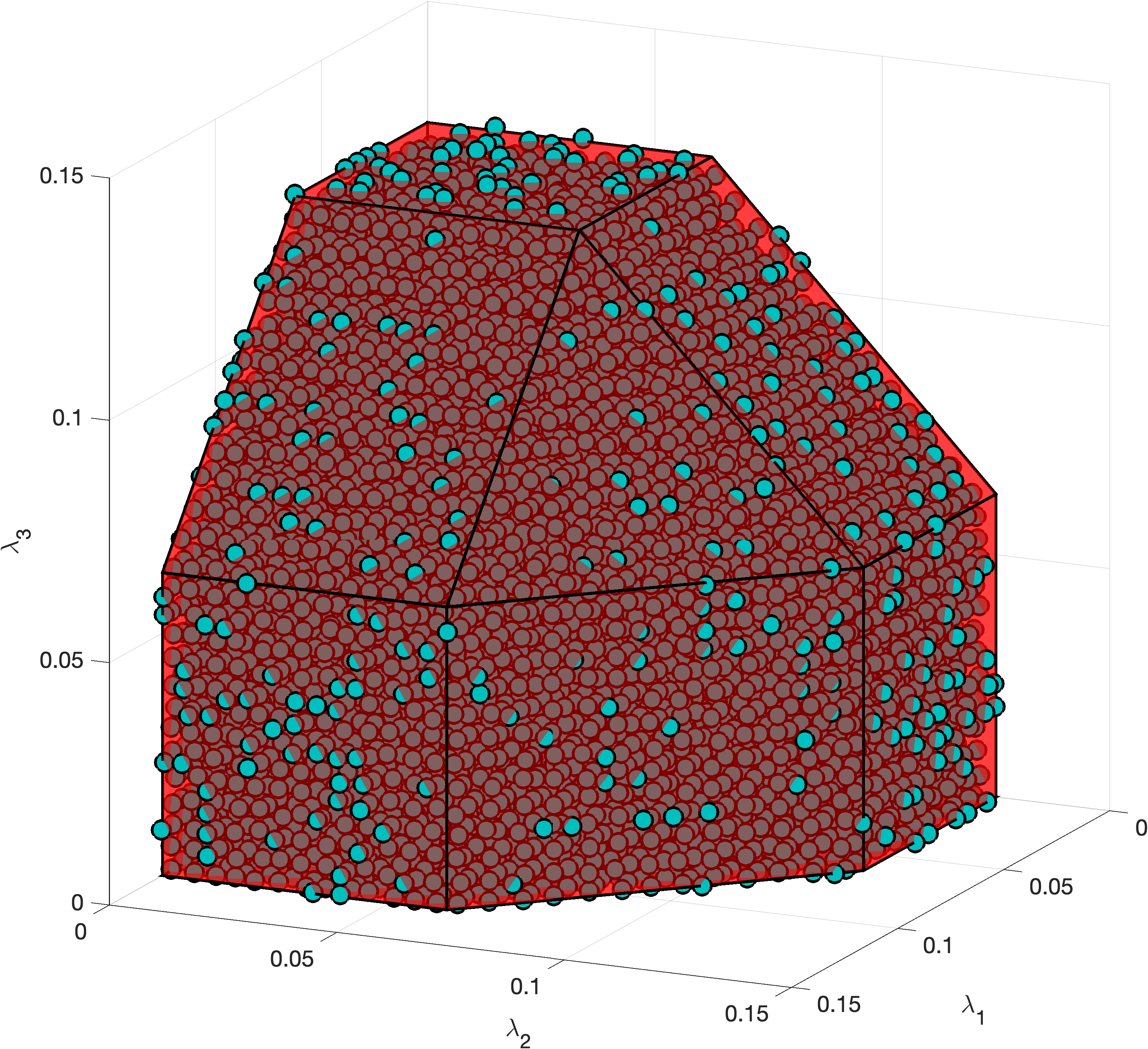}
    \caption{Servicing region $\Lambda^*$ for the triangular topology described in Section \ref{subsec:triangle1}.}
    \label{fig:triangleRegion}
\end{figure}

Recalling that all extreme points of $C$ must have exactly one nonzero value per column and that all elements of $C$ must be column stochastic, we recognize that we have $32 = 1\times 1\times 1 \times 2 \times 1 \times 2 \times 2 \times 4$ extreme points of the convex set of all $\mathbf{c}$ matrices.
We then perform the linear transformation $L(A) = \mathbf{Q}\bpi^* A \mathbf{P}$ on all of these points and remove the non-extreme points with an algorithm like QuickHull.
This leaves us with 13 unique extreme points of $\Lambda^*$, which can be further processed to give us 10 boundary planes of $\Lambda^*$.
Three of these planes simply enforce that $\lambda_i \geq 0$ for all $i$, so we say that there are $7$ non-trivial planes forming the boundary of $\Lambda^*$.
The region in question is shown in Fig. \ref{fig:triangleRegion} and analytically it can be expressed with the following planes (assume $i\neq j$ and $1\leq i,j \leq 3$):
\begin{align}
\label{eqn:triangleEqn}
    \frac{\lambda^*_i}{q_i} &\leq \frac{1}{4} \nonumber\\
	\frac{\lambda^*_i}{q_i} + \frac{\lambda^*_j}{q_j} &\leq \frac{3}{8} \nonumber\\
    \frac{\lambda^*_1}{q_1} + \frac{\lambda^*_2}{q_2} + \frac{\lambda^*_3}{q_3} &\leq \frac{1}{2} 
\end{align}
where we have generalized the swap probabilities to highlight specifically how they impact the geometry of the capacity region's boundaries.
This result can be understood with elementary combinatorial arguments that give necessary conditions on the capacity region~\cite{tillman2022continuous}, though this method of analysis does not readily scale up.

Now we go through the same calculation but for a CV switch.
This means we now have a $6$-flow model with $3^3=27$ possible elementary link arrangements, since each user's elementary link can be in three possible states (not existing, parity $1$, and parity $2$) and there are two ways to service each flow.
In the last section we defined the $\mathbf{a}^*$ matrix by having its rows count from $0$ to $2^K - 1$ in binary, and here we create a matrix that counts from $0$ to $3^K - 1$ in ternary.
Our $\bpi^*$ matrix simply expands to be a $6\times 6$ identity matrix with a column of all zeros appended on the left.
We also alter the $\mathbf{M}$ matrix by doubling the number of columns to account for there being twice as many flows, and doubling the number of rows to account for each user now having two different link parities:
$$
\mathbf{M}
=
\begin{bmatrix}
1 & 0 & 0 & 0 & 1 & 0 \\
0 & 1 & 0 & 0 & 0 & 1 \\
0 & 1 & 1 & 0 & 0 & 0 \\
1 & 0 & 0 & 1 & 0 & 0 \\
0 & 0 & 0 & 1 & 0 & 1 \\
0 & 0 & 1 & 0 & 1 & 0 \\
\end{bmatrix}.
$$
\noindent The first two rows correspond to user $1$, the next two correspond to user $2$, and the last two correspond to user $3$.
For example the first two columns correspond to flow $(1,2)$, the first column being when user $1$ has a parity $1$ elementary link while user 2's is parity $2$ and the second column corresponding to the opposite scenario.
All that is left to do is compute $\mathbb{P}(\mathbf{a}^{(j)})$ for every scenario $\mathbf{a}^{(j)}$, which is similar to the normal Bernoulli trial method except we need to enforce that one user cannot have both parities at the same time and we need to account for the fact that a successfully created elementary link has probability $1/2$ of being either parity.

\begin{figure}
    \centering
    \includegraphics[width=0.4\textwidth]{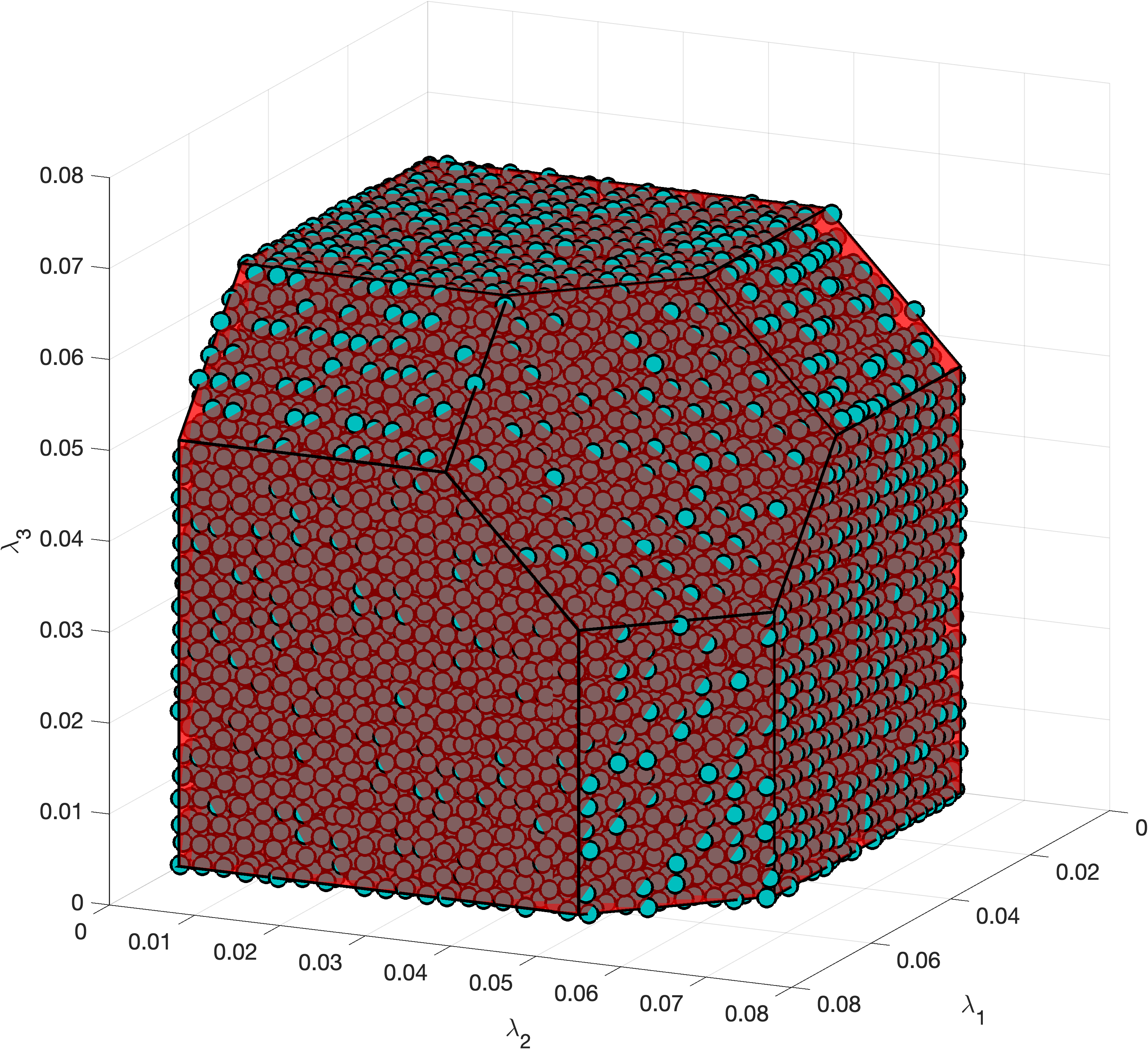}
    \caption{Servicing region $\Lambda^*$ for the triangular topology but with a CV protocol being used. Note that the maximum rate for any given flow is exactly half of the max rates shown in Fig. \ref{fig:triangleRegion}.}
    \label{fig:CV-triangle}
\end{figure}

Now that we redefined every necessary matrix, we can recompute $\mathcal{E}(C)$ and transform the set to get $450$ unique extreme points in $\mathbb{R}^6$.
If we want we can further reduce this to $16$ extreme points of $\mathbb{R}^3$ after combining virtual flows that correspond to the same real flow.
This 3-dimensional region is shown in Fig. \ref{fig:CV-triangle}.
Note that any specific flow has an upper bound of exactly $1/2$ of the upper bounds in Fig. \ref{fig:triangleRegion}, which makes sense because if two of the flows are almost entirely ignored then the only limiting factor on the third flow is the probability of success for both elementary links multiplied by the probability $1/2$ that the two links have opposite parity.

\begin{figure}
    \centering
    \includegraphics[width=0.4\textwidth]{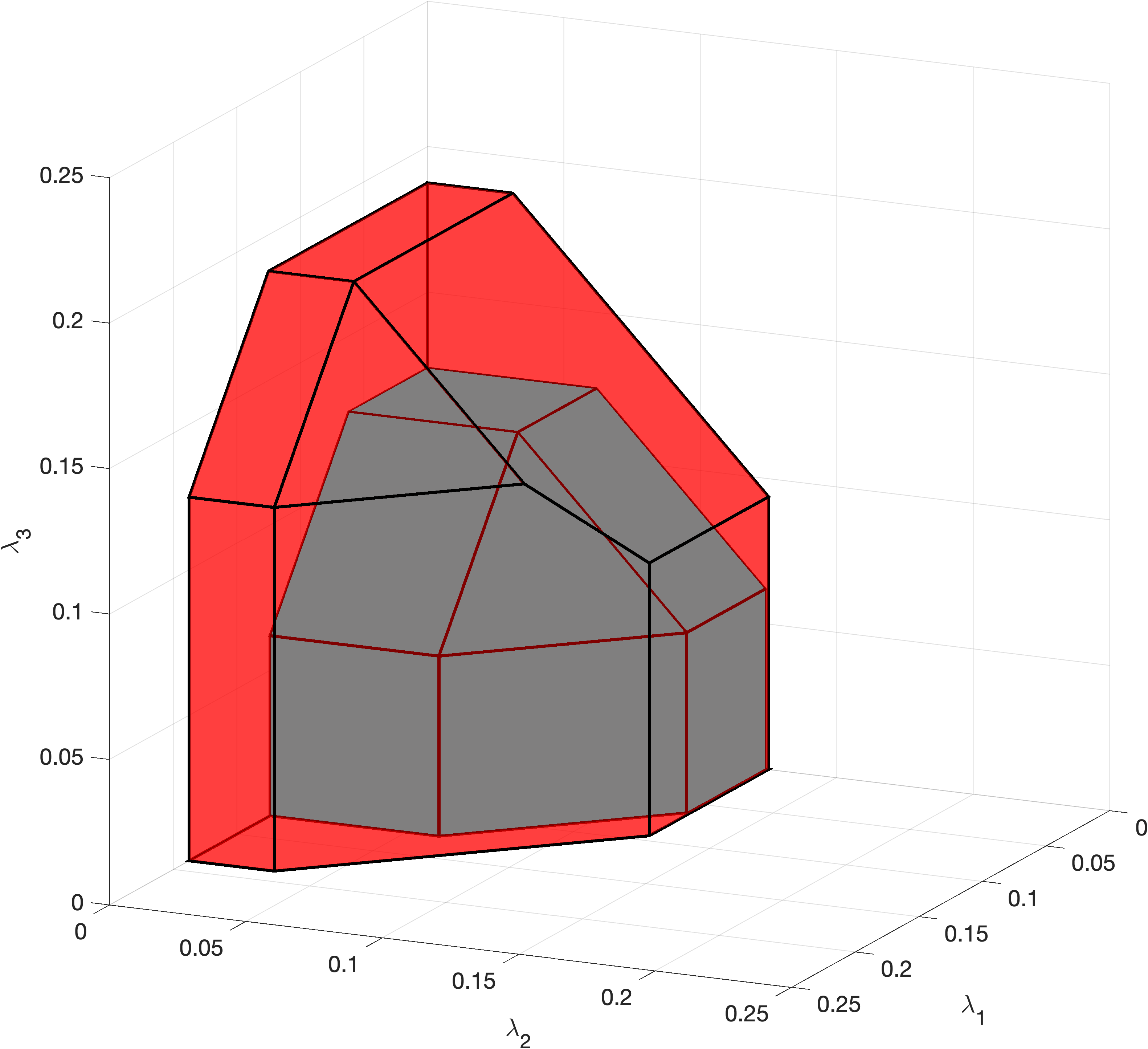}
    \caption{Servicing region $\Lambda^*$ for the triangular topology but where user $1$ is given two quantum memories instead of only one (red, larger). We include the capacity region of Fig. \ref{fig:triangleRegion} for reference (blue, smaller).}
    \label{fig:triangle211_comparison}
\end{figure}

\begin{figure}
    \centering
    \includegraphics[width=0.4\textwidth]{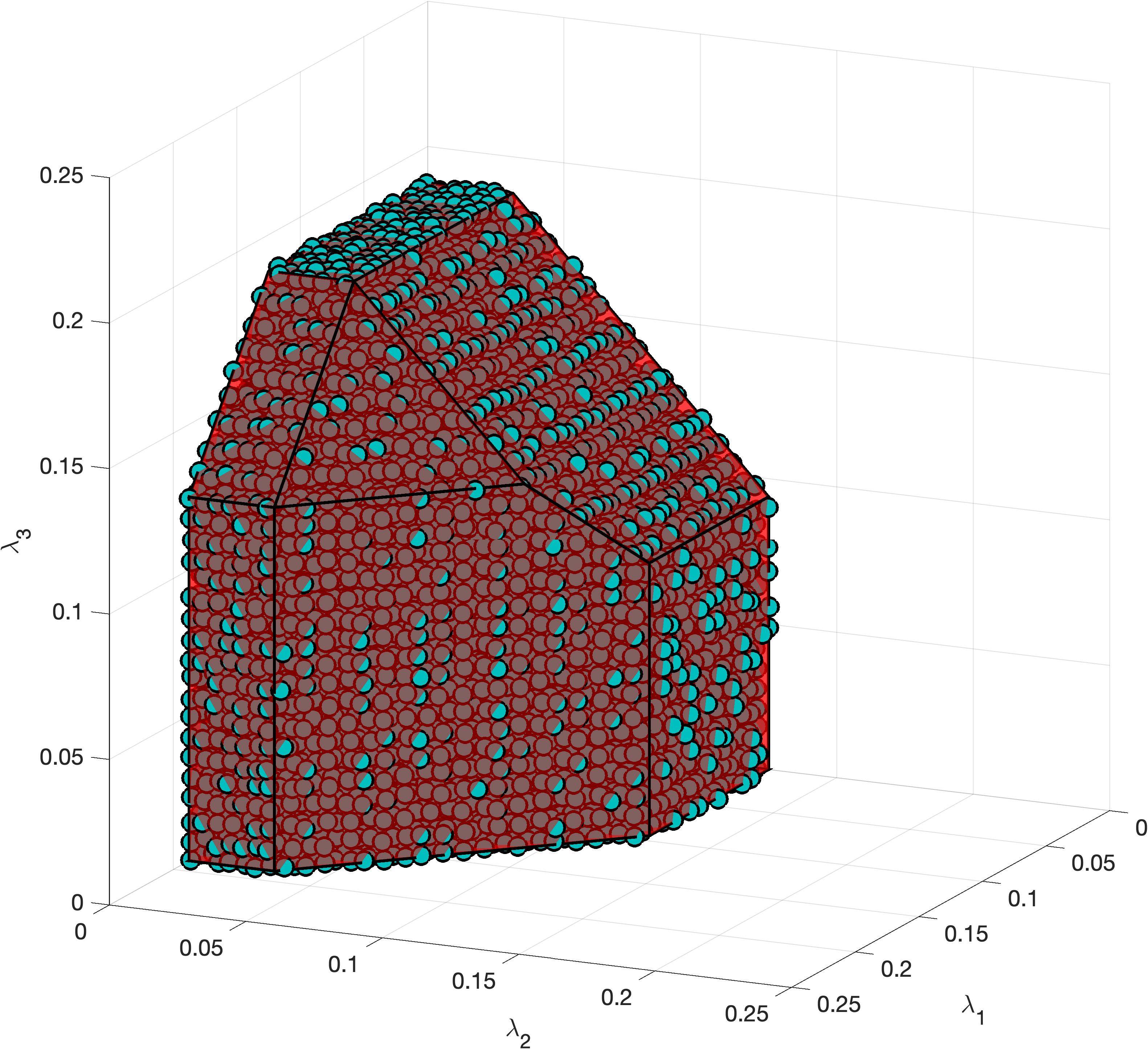}
    \caption{Servicing region $\Lambda^*$ for the triangular topology but where user $1$ is given two quantum memories with the simulated capacity region.}
    \label{fig:triangle211_alone}
\end{figure}

We analyze this triangular example one last time, now returning to DV and adding an extra memory to user $1$'s bank.
In other words, we set $D=\{ 2,1,1 \}$.
The new capacity region is shown in Fig. \ref{fig:triangle211_comparison} and Fig. \ref{fig:triangle211_alone}.
Clearly we have increased its size significantly, but it is interesting to note that the only new matching we are allowed to use is $\bpi = [1,0,1]^T$, so there is only a change in the region when considering flows $1$ and $3$.

\subsection{Square Topology}
\begin{table}[H]
    \normalsize
    \centering
    \setlength\extrarowheight{3pt}
    \begin{tabular}{c|c}
         Flows: & \{(1,2), (2,3), (3,4), (1,4)\}  \\
         $p$ & $\{ \frac{1}{2}, \frac{1}{2}, \frac{1}{2}, \frac{1}{2} \}$ \\
         $q$ & $\{ \frac{1}{2}, \frac{1}{2}, \frac{1}{2}, \frac{1}{2} \}$ \\
         $D$ & $\{1, 1, 1, 1\}$
    \end{tabular}
\end{table}

\begin{figure*}[ht]
     \centering
     \begin{subfigure}[b]{0.32\textwidth}
         \centering
         \includegraphics[width=\textwidth]{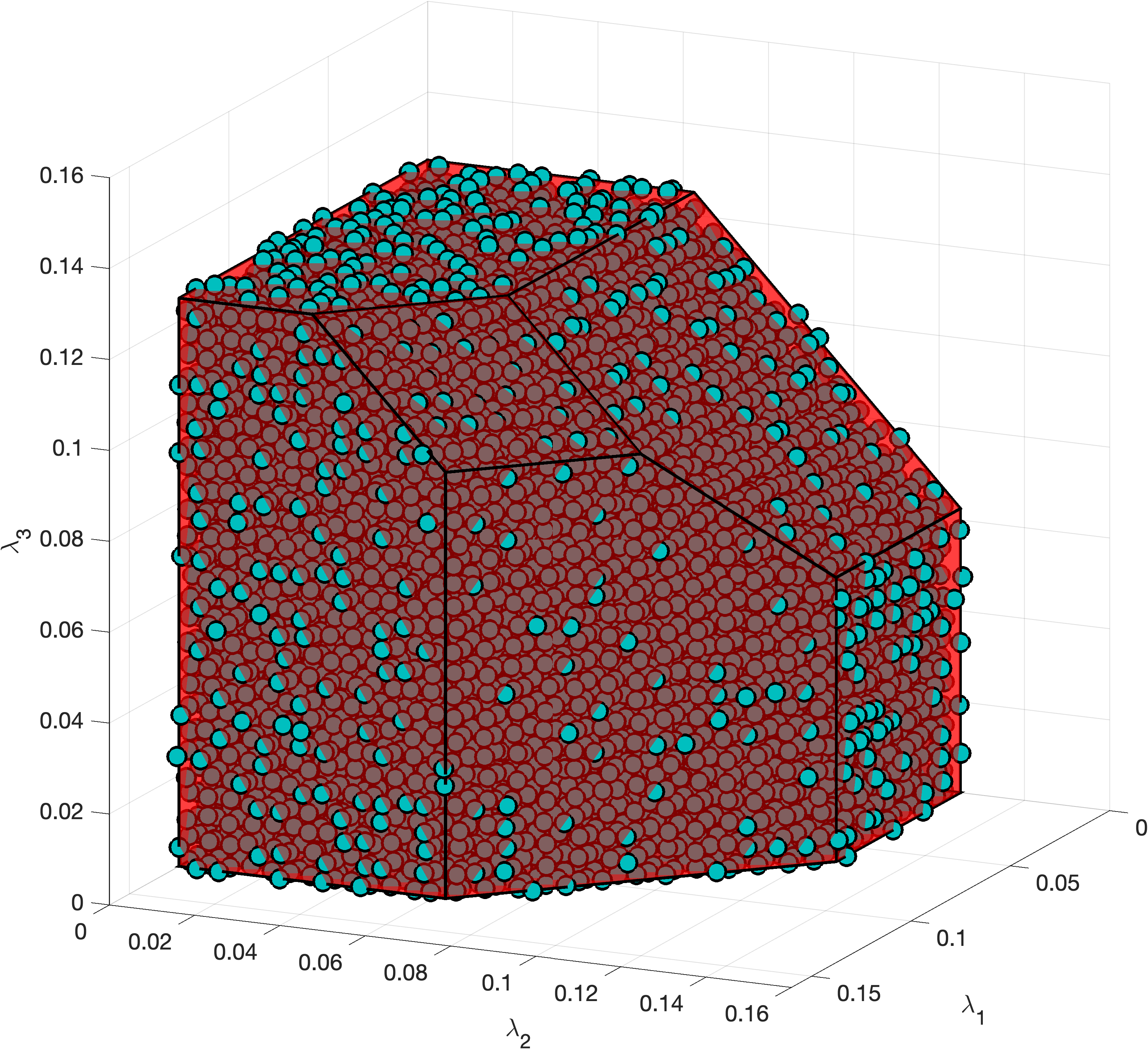}
         \caption{$\lambda_4=0$}
     \end{subfigure}
     \hfill
     \begin{subfigure}{0.32\textwidth}
         \centering
         \includegraphics[width=\textwidth]{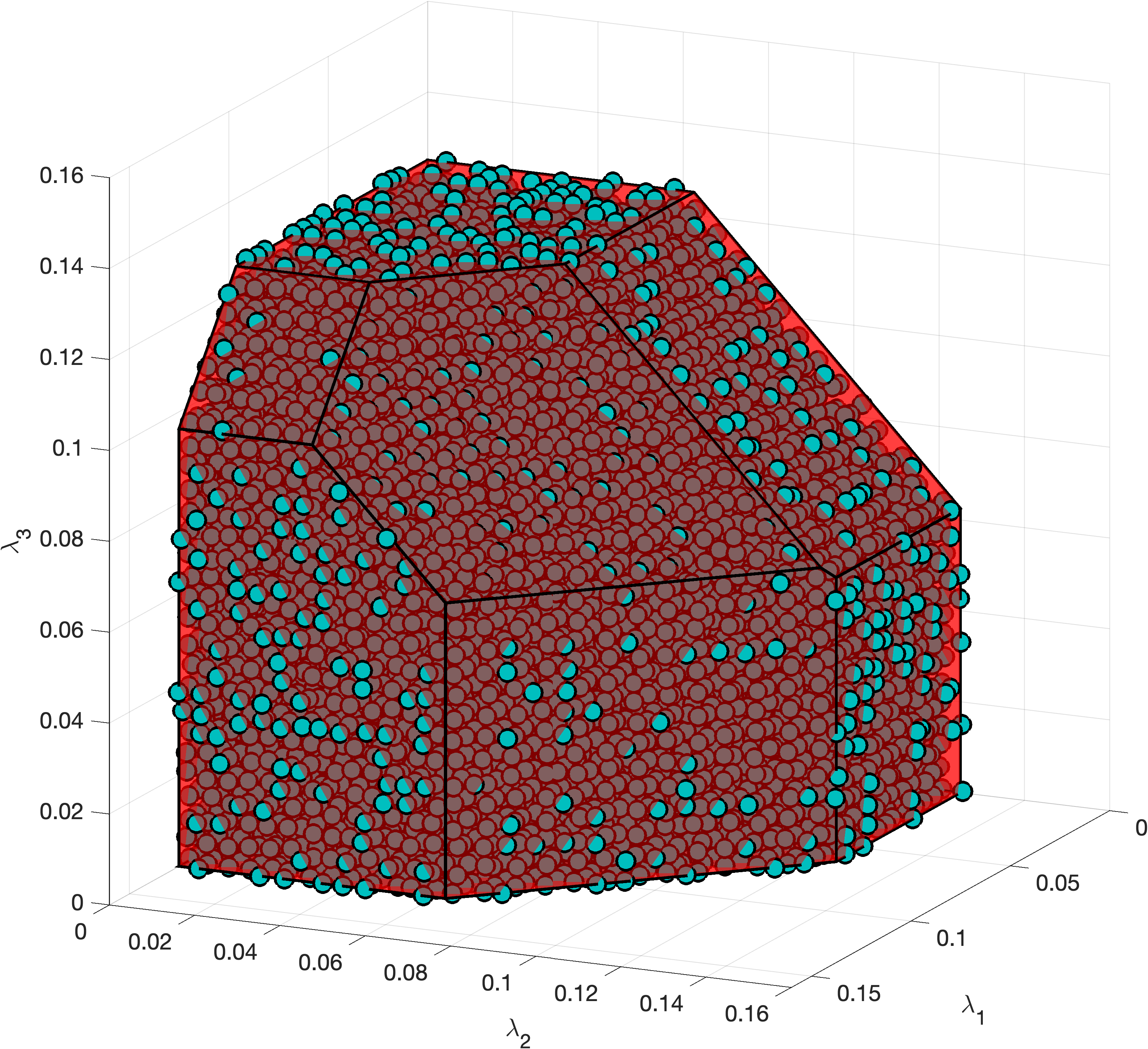}
         \caption{$\lambda_4=0.06$}
     \end{subfigure}
     \hfill
     \begin{subfigure}{0.32\textwidth}
         \centering
         \includegraphics[width=\textwidth]{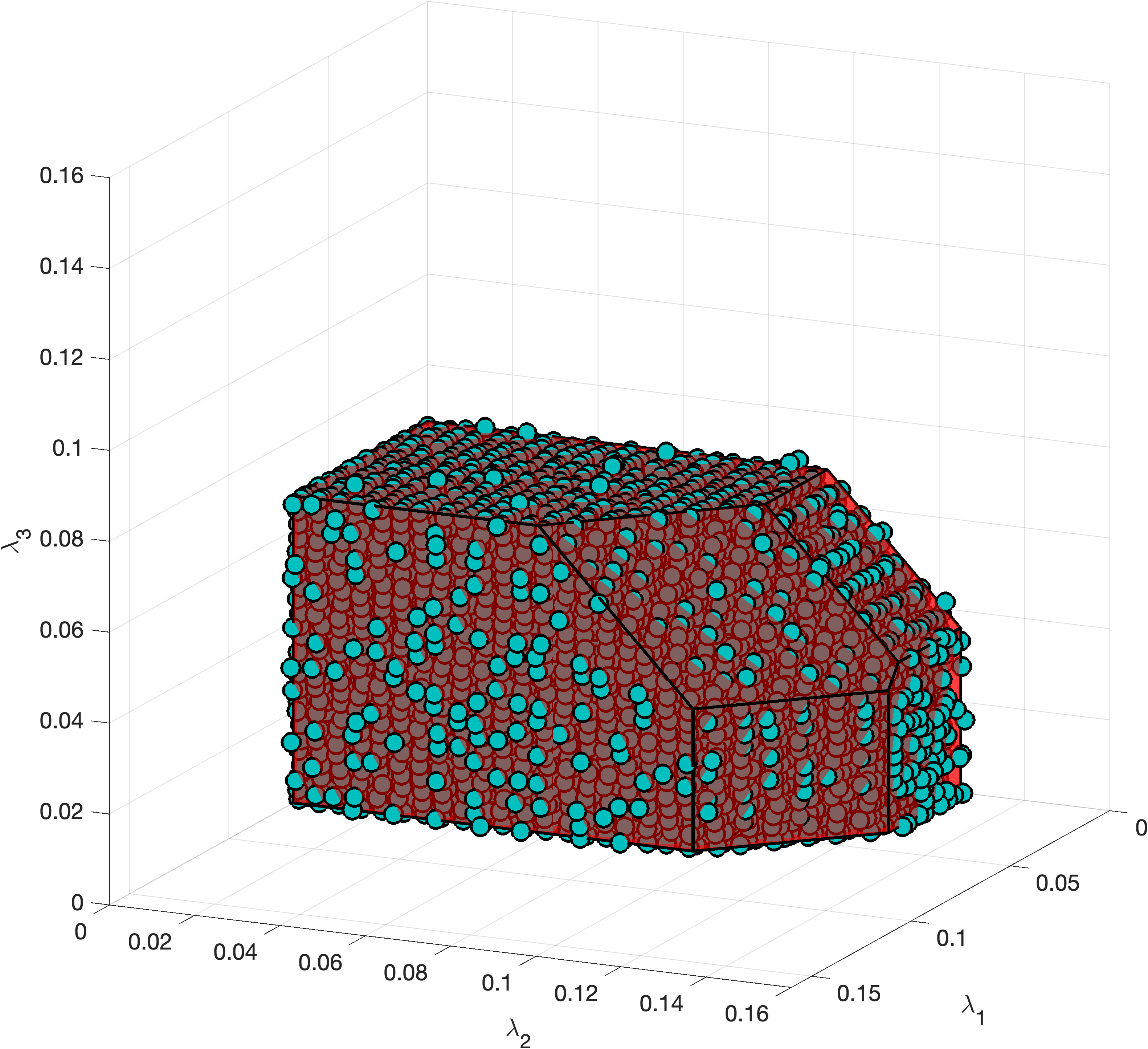}
         \caption{$\lambda_4=0.12$}
     \end{subfigure}
        \caption{Projections of the capacity region for a switch with four users connected to it, allowing any two adjacent users to request the generation of a bipartite end-to-end entanglement. We can see that these regions are symmetric about the plane $\lambda_1=\lambda_3$, while their extent on the $\lambda_2$ axis remains unchanged as we vary $\lambda_4$ because those flows do not directly compete.}
        \label{fig:squareCapacityRegion}
\end{figure*}

Just as in the last section we start with a polygon, place users with one memory at the vertices, and consider only the bipartite flows formed by the edges of the polygon.
An important change from the last example is that there are now scenarios where the switch can service multiple flows simultaneously (namely 1 and 3 or 2 and 4).
Because we are considering a four-flow model, we are no longer able to view the entire capacity region at once, so we show projections of $\Lambda^*$ for different values of $\lambda_4$.
These projections are shown in Fig. \ref{fig:squareCapacityRegion}.
The full region can again be expressed with the following planes (assume $i\neq j$, $i\neq k$, $j\neq k$, and $1\leq i,j,k\leq 4$):
\begin{align}
    \label{eqn:squareEqn}
    \frac{\lambda^*_i}{q_i} &\leq \frac{1}{4} \nonumber\\
	\frac{\lambda^*_i}{q_i} + \frac{\lambda^*_j}{q_j} &\leq \frac{3}{8} \nonumber\\
	\frac{\lambda^*_i}{q_i} + \frac{\lambda^*_j}{q_j} + \frac{\lambda^*_k}{q_k} &\leq \frac{9}{16} \nonumber\\
    \frac{\lambda^*_1}{q_1} + \frac{\lambda^*_2}{q_2} + \frac{\lambda^*_3}{q_3} + \frac{\lambda^*_4}{q_4} &\leq \frac{5}{8}
\end{align}

We unfortunately cannot compute the capacity region for the CV case of the square model because moving to an $8$-flow model increases $|\mathcal{E}(C)|$ to almost $10^{19}$, an infeasible size for brute force methods.

\subsection{Pentagon with mixed-partite flows}
\label{subsec:pentagon}
\begin{figure*}[ht]
     \centering
     \begin{subfigure}[b]{0.3\textwidth}
         \centering
         \includegraphics[width=\textwidth]{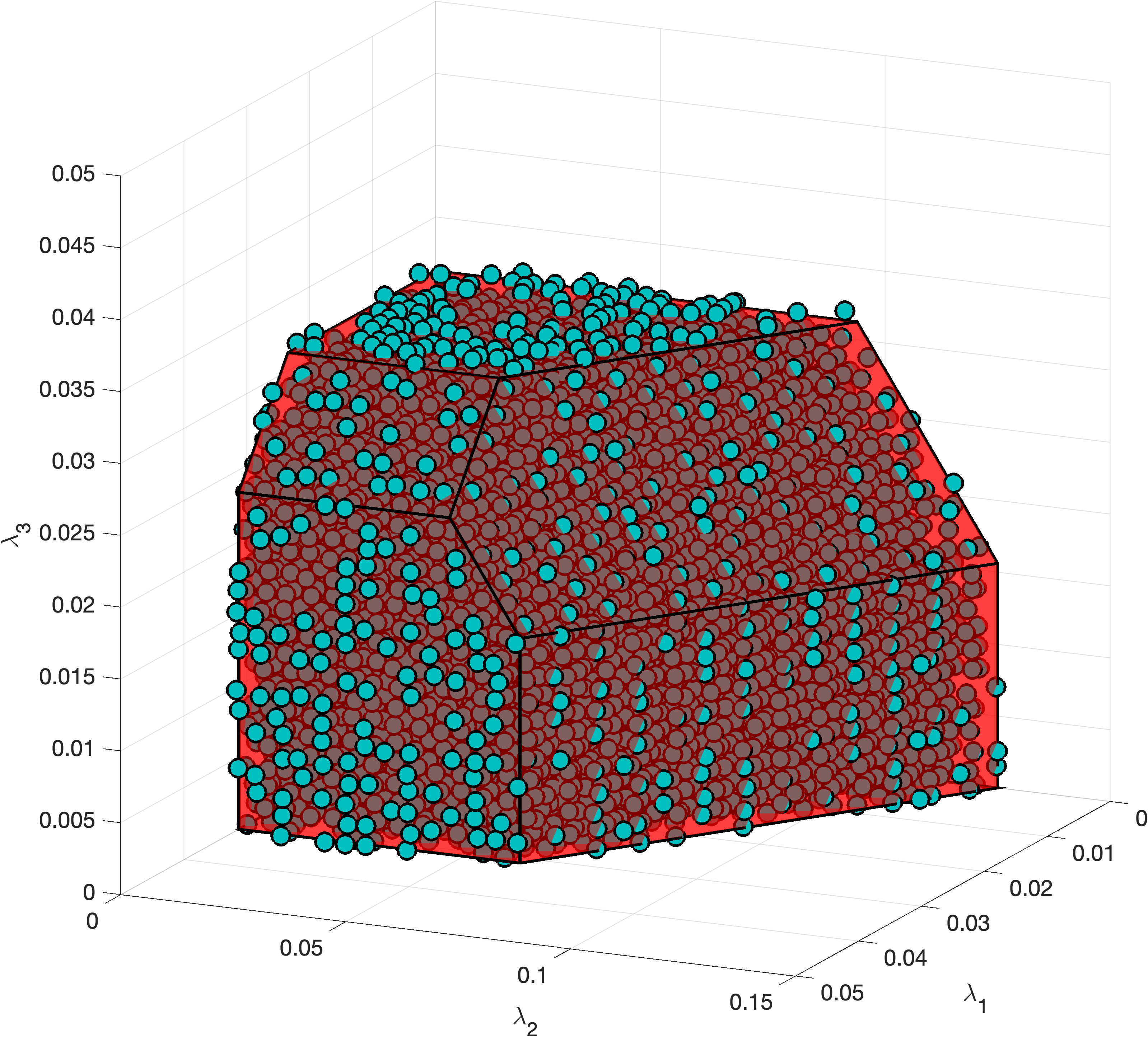}
         \caption{$\lambda_4=0$, $\lambda_5=0$}
     \end{subfigure}
     \hfill
     \begin{subfigure}{0.32\textwidth}
         \centering
         \includegraphics[width=\textwidth]{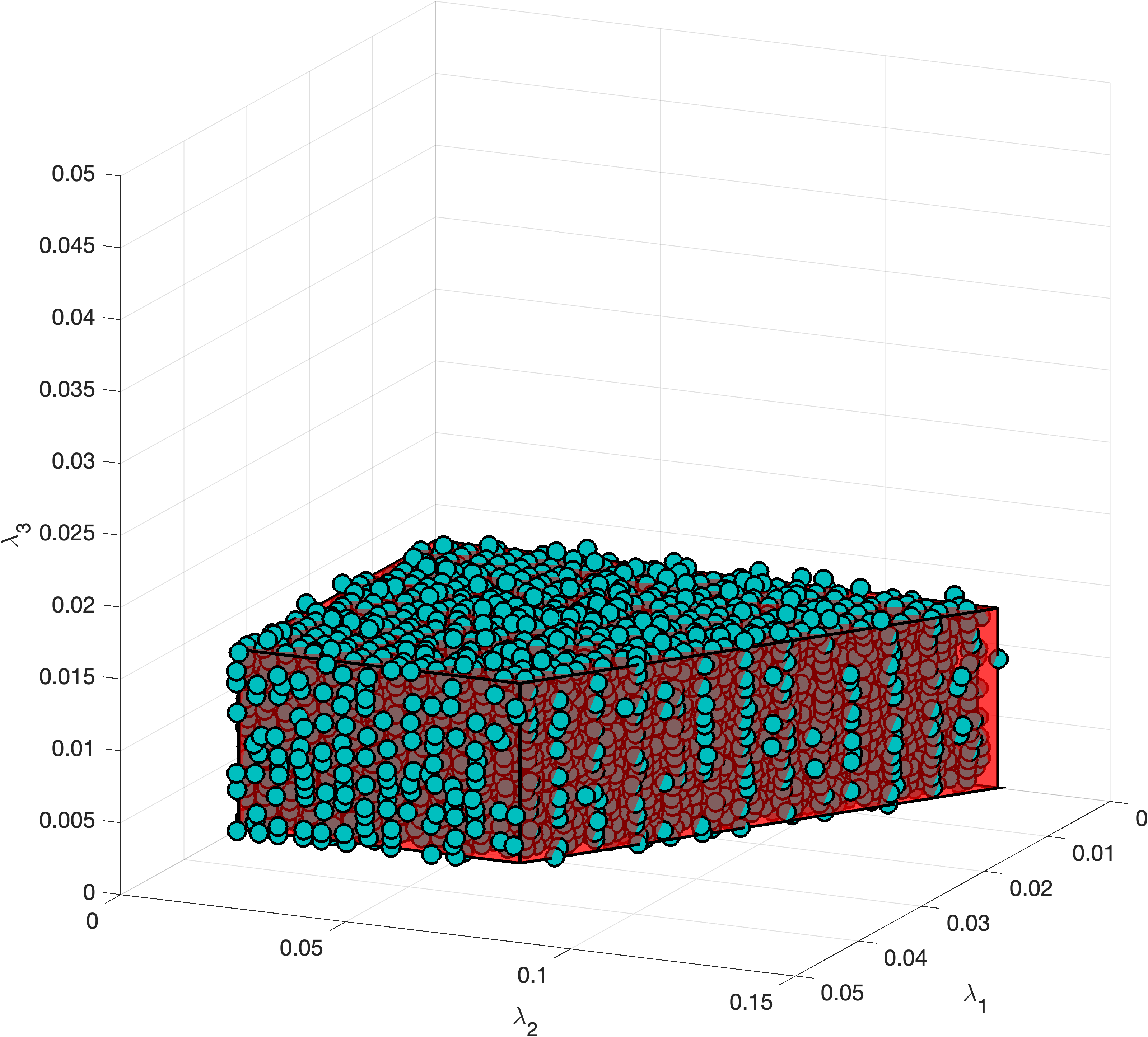}
         \caption{$\lambda_4=0.1$, $\lambda_5=0$}
     \end{subfigure}
     \hfill
     \begin{subfigure}{0.3\textwidth}
         \centering
         \includegraphics[width=\textwidth]{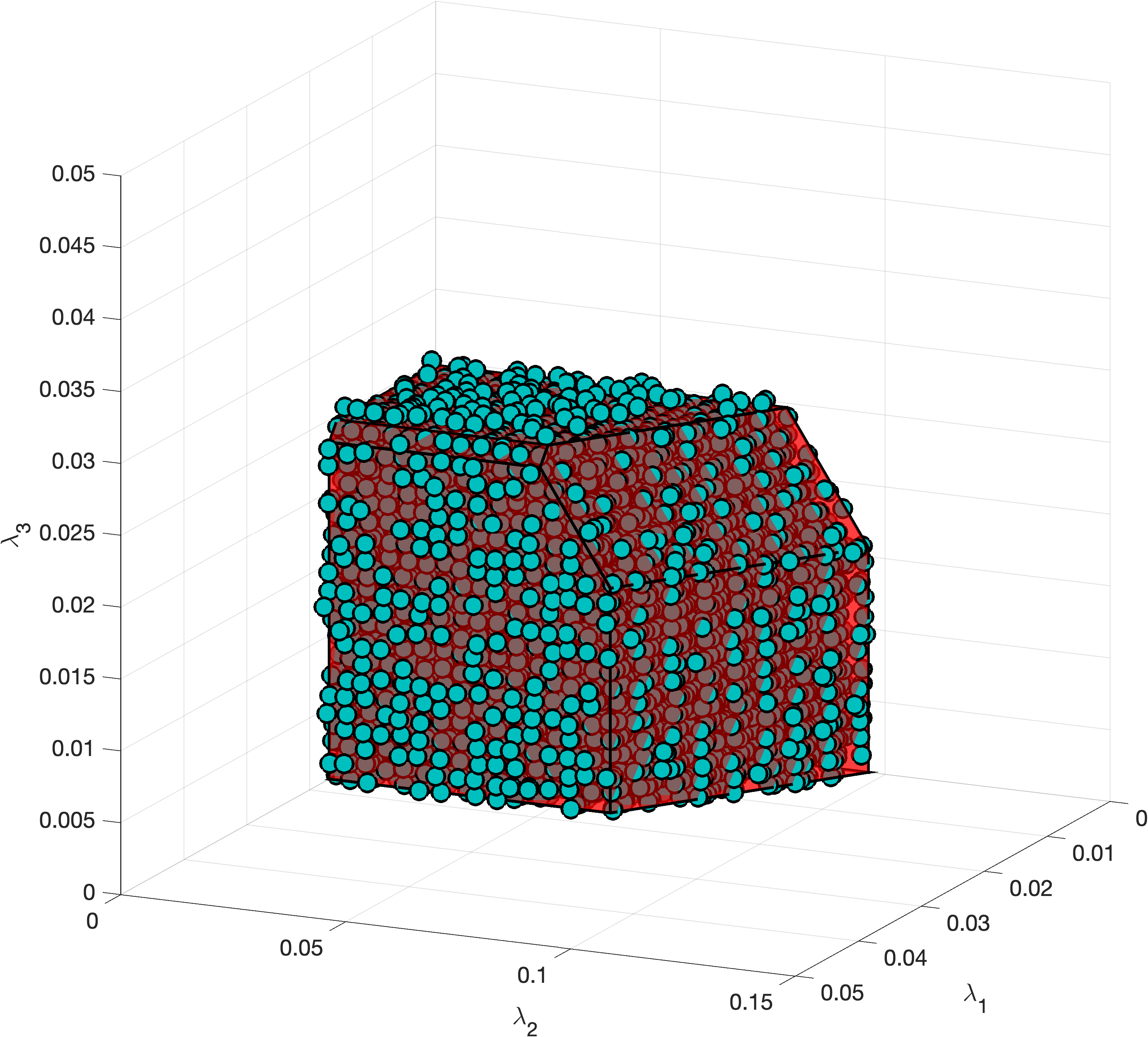}
         \caption{$\lambda_4=0$, $\lambda_5=0.03$}
     \end{subfigure}
        \caption{Projections of the capacity region for a switch with five users connected to it, where we only consider the flows named in Section \ref{subsec:pentagon}. By increasing the servicing rate for either flow 4 or flow 5 we can decrease the possible servicing rates for the flows they compete with.}
        \label{fig:pentagonCapacityRegion}
\end{figure*}

Looking at (\ref{eqn:triangleEqn}) and (\ref{eqn:squareEqn}) one might conjecture that we only need bounds of the form
\begin{equation}
\label{eqn:subsetSum}
\sum_{k\in W}\frac{\lambda^*_k}{q_k} \leq \beta_W
\end{equation}
for various subsets $W\subseteq \{1,\cdots,M\}$, but here we show an example where we need to weight the flows differently.
Consider a more complicated example with both bipartite and tripartite flows:
\begin{table}[H]
    \normalsize
    \centering
    \setlength\extrarowheight{3pt}
    \begin{tabular}{c|c}
         Flows: & \{(1,2,3), (2,3), (3,4,5), (4,5), (5,1,2)\}  \\
         $p$ & $\{ \frac{1}{2}, \frac{1}{2}, \frac{1}{2}, \frac{1}{2}, \frac{1}{2} \}$ \\
         $q$ & $\{ \frac{1}{4}, \frac{1}{2}, \frac{1}{4}, \frac{1}{2}, \frac{1}{4} \}$ \\
         $D$ & $\{1, 1, 1, 1, 1\}$
    \end{tabular}
\end{table}
\noindent We consider this model both to increase the complexity of the competition between flows and to avoid the infeasible calculation of the pentagon of users with bipartite flows.
Note that we have reduced the swap probabilities for the tripartite flows.
As a reminder, the specific probabilities that we use are not based on any physical model -- they can be replaced with any probabilities that come from specific models of entanglement generation and entanglement swapping.
Our $\bpi^*$ matrix has a slightly more complicated form due to the irregular overlaps of the different flows:
$$
\bpi^*
=
\begin{bmatrix}
0 & 1 & 0 & 0 & 0 & 0 & 1 & 0 \\
0 & 0 & 1 & 0 & 0 & 0 & 0 & 1 \\
0 & 0 & 0 & 1 & 0 & 0 & 0 & 0 \\
0 & 0 & 0 & 0 & 1 & 0 & 1 & 1 \\
0 & 0 & 0 & 0 & 0 & 1 & 0 & 0 \\
\end{bmatrix}.
$$
The $\mathbf{c}$ matrix will now be $8 \times 32$ because we have $8$ flows and $2^5=32$ possible arrangements of the elementary links.
After constructing this matrix by setting $c_{i,j}\rightarrow 0$ if $M\bpi^{(i)} \nleq \mathbf{a}^{(j)} $ and doing some arithmetic, we see that $|\mathcal{E}\left( C \right)| = 4976640$ which is a reasonable number of extreme points to brute force transform one-by-one.
After performing $Q\bpi^* \mathcal{E}(C)\mathbf{P}$ we get a set with $4448$ unique points, $105$ of which are extreme points forming 22 boundary planes (17 non-trivial) of the servicing region $\Lambda^*$.
We show projections of this region in Fig. \ref{fig:pentagonCapacityRegion}.
Along with inequalities of the form shown in (\ref{eqn:subsetSum}), which we will omit for brevity, we also need the following two inequalities to describe the boundary of the capacity region:
\begin{align*}
    \frac{\lambda^*_1}{q_1} + 2\frac{\lambda^*_3}{q_3} + \frac{\lambda^*_4}{q_4} + \frac{\lambda^*_5}{q_5} &\leq  \frac{1}{2} \\
    \frac{\lambda^*_1}{q_1} + \frac{\lambda^*_2}{q_2} + 2\frac{\lambda^*_3}{q_3} + \frac{\lambda^*_4}{q_4} + \frac{\lambda^*_5}{q_5} &\leq  \frac{19}{32}
\end{align*}
which require non-unity coefficients on the rates.

We choose not to analyze the CV version of this network because, using our model for CV state generation, the optimal end-to-end state generation protocol is only known for bipartite links~\cite{tillman2022supporting}.
It is an open question to find the optimal arrangement of elementary link parities when creating a multipartite entangled state with a swap operation acting on NLA- and TMSV-based links.

\section{Discussion \label{sec:discussion}}
\subsection{Shortcut for a Particular Use Case}
\label{subsec:shortcuts}
A problem one might be interested in solving is maximizing a weighted sum of servicing rates over the servicing region, i.e. calculating
\begin{equation}
\label{eqn:metric}
     \max_{\blam^* \in \Lambda^*}\sum_j \alpha_j \lambda^*_j. 
\end{equation}
One option to solve this is to use Section \ref{sec:results} to calculate the capacity region and then maximize over its boundaries, but it turns out this specific problem can be solved much more easily with a different method.
This can be done by applying elementary linear programming directly to (\ref{eqn:servicingRate}) instead of calculating the entire capacity region, which is the computationally expensive part behind all results previously shown.
This shortcut could be very useful to a network administrator that, instead of taking in requests for entanglement, just generates whatever end-to-end entangled state they choose in a given timestep. 
They will then most likely have a metric that they are trying to optimize, and many reasonable metrics will take the form of (\ref{eqn:metric}).

\subsection{Computational Complexity}
Calculating $\mathcal{E}(C)$ for even modest topologies can be infeasible.
For example if we look at a topology where users are vertices on a regular $K$-sided polygon and we only consider the bipartite entanglement flows between adjacent users, i.e. the edges of the polygon, then $|\mathcal{E}(C_K)|$ grows faster than $2^{2^K}$ where we define $C_K$ as the set of valid scheduling matrices for the $K$-sided polygon topology.
Expanding upon Section \ref{subsec:pentagon}, analyzing the pentagonal case ($K=5$) requires transforming almost $10^{10}$ extreme points of $C$ if the proposed algorithm is followed strictly.
There are almost certainly improvements that can be made, but the goal of this paper was to prove that a general strategy exists.

\section{Conclusion}
In summary, we presented an algorithm based on tools from convex analysis for calculating the boundary of the capacity region for a bufferless quantum switch.
While previous work had successfully characterized the capacity region, there was no immediately obvious way of obtaining its boundaries from their results.
Afterwards we presented several example calculations using our algorithm, some that can also be calculated by simple combinatorial methods and some that cannot.

Some obvious ways to extend this work include allowing elementary links to survive for multiple timesteps (i.e. $\tau > 1$), switching to a continuous time model rather than using slotted time, and including a purification protocol after many end-to-end states are generated.
One could also work on improving the efficiency of the algorithm or finding shortcuts, such as proving that all boundary planes will be of a certain form and then using Section \ref{subsec:shortcuts} to jump straight to the boundaries of $\Lambda^*$.
We will also use this section to reiterate that the capacity region, $\Lambda$, is simply the servicing region minus its boundary, so although all calculations shown have been used to calculate the servicing region, $\Lambda^*$, one can trivially get the capacity region from the results shown.

\bibliography{references.bib}

\begin{thebibliography}{10}

\bibitem{kleinrock2010early}
Leonard Kleinrock.
\newblock An early history of the internet [history of communications].
\newblock {\em IEEE Communications Magazine}, 48(8):26--36, 2010.

\bibitem{hui2019introduction}
Rongqing Hui.
\newblock {\em Introduction to fiber-optic communications}.
\newblock Academic Press, 2019.

\bibitem{bertsekas2021data}
Dimitri Bertsekas and Robert Gallager.
\newblock {\em Data networks}.
\newblock Athena Scientific, 2021.

\bibitem{WEH18}
Stephanie Wehner, David Elkouss, and Ronald Hanson.
\newblock Quantum internet: A vision for the road ahead.
\newblock {\em Science}, 362(6412):eaam9288, 2018.

\bibitem{dowlingMilburnQuantumRevolution}
Jonathan~P Dowling and Gerard~J Milburn.
\newblock Quantum technology: the second quantum revolution.
\newblock {\em Philosophical Transactions of the Royal Society of London.
  Series A: Mathematical, Physical and Engineering Sciences},
  361(1809):1655--1674, 2003.

\bibitem{kimbleQuantumInternet}
H~Jeff Kimble.
\newblock The quantum internet.
\newblock {\em Nature}, 453(7198):1023--1030, 2008.

\bibitem{brady2022entangled}
Anthony~J Brady, Christina Gao, Roni Harnik, Zhen Liu, Zheshen Zhang, and
  Quntao Zhuang.
\newblock Entangled sensor-networks for dark-matter searches.
\newblock {\em PRX Quantum}, 3(3):030333, 2022.

\bibitem{quantumSensingReview}
Stefano Pirandola, B~Roy Bardhan, Tobias Gehring, Christian Weedbrook, and Seth
  Lloyd.
\newblock Advances in photonic quantum sensing.
\newblock {\em Nature Photonics}, 12(12):724--733, 2018.

\bibitem{diadamo2022packet}
Stephen DiAdamo, Bing Qi, Glen Miller, Ramana Kompella, and Alireza Shabani.
\newblock Packet switching in quantum networks: A path to the quantum internet.
\newblock {\em Physical Review Research}, 4(4):043064, 2022.

\bibitem{dai2021entanglement}
Wenhan Dai, Anthony Rinaldi, and Don Towsley.
\newblock Entanglement swapping in quantum switches: Protocol design and
  stability analysis.
\newblock {\em arXiv preprint arXiv:2110.04116}, 2021.

\bibitem{vasantam2021stability}
Thirupathaiah Vasantam and Don Towsley.
\newblock Stability analysis of a quantum network with max-weight scheduling.
\newblock {\em arXiv preprint arXiv:2106.00831}, 2021.

\bibitem{chung2022design}
Joaquin Chung, Ely~M Eastman, Gregory~S Kanter, Keshav Kapoor, Nikolai Lauk,
  Cristian~H Pena, Robert~K Plunkett, Neil Sinclair, Jordan~M Thomas, Raju
  Valivarthi, et~al.
\newblock Design and implementation of the illinois express quantum
  metropolitan area network.
\newblock {\em IEEE Transactions on Quantum Engineering}, 3:1--20, 2022.

\bibitem{clark2023entanglement}
Marcus~J Clark, Obada Alia, Rui Wang, Sima Bahrani, M~Perani{\'c}, Djeylan
  Aktas, George~T Kanellos, Martin Loncaric, {\v{Z}}~Samec, Anton Radman,
  et~al.
\newblock Entanglement distribution quantum networking within deployed
  telecommunications fibre-optic infrastructure.
\newblock In {\em Quantum Technology: Driving Commercialisation of an Enabling
  Science III}, volume 12335, pages 96--103. SPIE, 2023.

\bibitem{harney2022analytical}
Cillian Harney and Stefano Pirandola.
\newblock Analytical methods for high-rate global quantum networks.
\newblock {\em PRX Quantum}, 3(1):010349, 2022.

\bibitem{chen2023zero}
Kevin~C Chen, Prajit Dhara, Mikkel Heuck, Yuan Lee, Wenhan Dai, Saikat Guha,
  and Dirk Englund.
\newblock Zero-added-loss entangled-photon multiplexing for ground-and
  space-based quantum networks.
\newblock {\em Physical Review Applied}, 19(5):054029, 2023.

\bibitem{cicconetti2022resource}
Claudio Cicconetti, Marco Conti, and Andrea Passarella.
\newblock Resource allocation in quantum networks for distributed quantum
  computing.
\newblock In {\em 2022 IEEE International Conference on Smart Computing
  (SMARTCOMP)}, pages 124--132. IEEE, 2022.

\bibitem{thiru}
Thirupathaiah Vasantam and Don Towsley.
\newblock {A throughput optimal scheduling policy for a quantum switch}.
\newblock In Philip~R. Hemmer and Alan~L. Migdall, editors, {\em Quantum
  Computing, Communication, and Simulation II}, volume 12015, pages 14 -- 23.
  International Society for Optics and Photonics, SPIE, 2022.

\bibitem{VGNT21}
Gayane Vardoyan, Saikat Guha, Philippe Nain, and Don Towsley.
\newblock On the stochastic analysis of a quantum entanglement distribution
  switch.
\newblock {\em IEEE Transactions on Quantum Engineering}, 2:1--16, 2021.

\bibitem{vardoyan2023capacity}
Gayane Vardoyan, Philippe Nain, Saikat Guha, and Don Towsley.
\newblock On the capacity region of bipartite and tripartite entanglement
  switching.
\newblock {\em ACM Transactions on Modeling and Performance Evaluation of
  Computing Systems}, 8(1-2):1--18, 2023.

\bibitem{promponas2023full}
Panagiotis Promponas, V{\'\i}ctor Valls, and Leandros Tassiulas.
\newblock Full exploitation of limited memory in quantum entanglement
  switching.
\newblock {\em arXiv preprint arXiv:2304.10602}, 2023.

\bibitem{tillman2022continuous}
Ian Tillman, Thirupathaiah Vasantam, and Kaushik~P Seshadreesan.
\newblock A continuous variable quantum switch.
\newblock In {\em 2022 IEEE International Conference on Quantum Computing and
  Engineering (QCE)}, pages 365--371. IEEE, 2022.

\bibitem{noCloning}
William~K Wootters and Wojciech~H Zurek.
\newblock A single quantum cannot be cloned.
\newblock {\em Nature}, 299(5886):802--803, 1982.

\bibitem{PLOB}
Stefano Pirandola, Riccardo Laurenza, Carlo Ottaviani, and Leonardo Banchi.
\newblock Fundamental limits of repeaterless quantum communications.
\newblock {\em Nature communications}, 8(1):1--15, 2017.

\bibitem{munro2015inside}
William~J Munro, Koji Azuma, Kiyoshi Tamaki, and Kae Nemoto.
\newblock Inside quantum repeaters.
\newblock {\em IEEE Journal of Selected Topics in Quantum Electronics},
  21(3):78--90, 2015.

\bibitem{repeaterGenerations}
Sreraman Muralidharan, Linshu Li, Jungsang Kim, Norbert L{\"u}tkenhaus,
  Mikhail~D Lukin, and Liang Jiang.
\newblock Optimal architectures for long distance quantum communication.
\newblock {\em Scientific reports}, 6(1):1--10, 2016.

\bibitem{yan2021survey}
Pei-Shun Yan, Lan Zhou, Wei Zhong, and Yu-Bo Sheng.
\newblock A survey on advances of quantum repeater.
\newblock {\em Europhysics Letters}, 136(1):14001, 2021.

\bibitem{azuma2022quantum}
Koji Azuma, Sophia~E Economou, David Elkouss, Paul Hilaire, Liang Jiang,
  Hoi-Kwong Lo, and Ilan Tzitrin.
\newblock Quantum repeaters: From quantum networks to the quantum internet.
\newblock {\em arXiv preprint arXiv:2212.10820}, 2022.

\bibitem{ralph2009nondeterministic}
Timothy~C Ralph and AP~Lund.
\newblock Nondeterministic noiseless linear amplification of quantum systems.
\newblock In {\em AIP Conference Proceedings}, pages 155--160. American
  Institute of Physics, 2009.

\bibitem{he2021noiseless}
Mingjian He, Robert Malaney, and Benjamin~A Burnett.
\newblock Noiseless linear amplifiers for multimode states.
\newblock {\em Physical Review A}, 103(1):012414, 2021.

\bibitem{PJCC13}
Shashank Pandey, Zhang Jiang, Joshua Combes, and Carlton~M. Caves.
\newblock Quantum limits on probabilistic amplifiers.
\newblock {\em Phys. Rev. A}, 88:033852, Sep 2013.

\bibitem{peggQuantumScissor}
David~T Pegg, Lee~S Phillips, and Stephen~M Barnett.
\newblock Optical state truncation by projection synthesis.
\newblock {\em Physical review letters}, 81(8):1604, 1998.

\bibitem{diasRalph2018}
Josephine Dias and Timothy~C Ralph.
\newblock Quantum error correction of continuous-variable states with realistic
  resources.
\newblock {\em Physical Review A}, 97(3):032335, 2018.

\bibitem{kaushikPRR}
Kaushik~P Seshadreesan, Hari Krovi, and Saikat Guha.
\newblock Continuous-variable quantum repeater based on quantum scissors and
  mode multiplexing.
\newblock {\em Physical Review Research}, 2(1):013310, 2020.

\bibitem{DLCZ}
L-M Duan, Mikhail~D Lukin, J~Ignacio Cirac, and Peter Zoller.
\newblock Long-distance quantum communication with atomic ensembles and linear
  optics.
\newblock {\em Nature}, 414(6862):413--418, 2001.

\bibitem{tillman2022supporting}
Ian~J Tillman, Allison Rubenok, Saikat Guha, and Kaushik~P Seshadreesan.
\newblock Supporting multiple entanglement flows through a continuous-variable
  quantum repeater.
\newblock {\em Physical Review A}, 106(6):062611, 2022.

\bibitem{diasCVRepeater}
Josephine Dias, Matthew~S Winnel, Nedasadat Hosseinidehaj, and Timothy~C Ralph.
\newblock Quantum repeater for continuous-variable entanglement distribution.
\newblock {\em Physical Review A}, 102(5):052425, 2020.

\bibitem{panigrahy2023capacity}
Nitish~K Panigrahy, Thirupathaiah Vasantam, Don Towsley, and Leandros
  Tassiulas.
\newblock On the capacity region of a quantum switch with entanglement
  purification.
\newblock In {\em IEEE INFOCOM 2023-IEEE Conference on Computer
  Communications}, pages 1--10. IEEE, 2023.

\bibitem{simon2011convexity}
Barry Simon.
\newblock {\em Convexity: an analytic viewpoint}, volume 187.
\newblock Cambridge University Press, 2011.

\bibitem{quickhull}
C~Bradford Barber, David~P Dobkin, and Hannu Huhdanpaa.
\newblock The quickhull algorithm for convex hulls.
\newblock {\em ACM Transactions on Mathematical Software (TOMS)},
  22(4):469--483, 1996.

\bibitem{preparata2012computational}
Franco~P Preparata and Michael~I Shamos.
\newblock {\em Computational geometry: an introduction}.
\newblock Springer Science \& Business Media, 2012.

\bibitem{furrerAndMunro}
Fabian Furrer and William~J Munro.
\newblock Repeaters for continuous-variable quantum communication.
\newblock {\em Physical Review A}, 98(3):032335, 2018.

\end{thebibliography}
\bibliographystyle{unsrt}

\end{document}